\documentclass[prc,twocolumn,showpacs,preprintnumbers,superscriptaddress,floatfix,nofootinbib]{revtex4-1}
\usepackage{amssymb}
\usepackage{amsmath,txfonts}
\usepackage{graphicx}
\usepackage{dcolumn}
\usepackage{color}
\usepackage{bm}
\usepackage[subfigure]{graphfig}
\usepackage{makecell}

\begin{document}

\title{\boldmath The $K^{\ast 0}\Lambda$ photoproduction off a neutron}
\author{Xiao-Yun Wang}
\thanks{xywang@impcas.ac.cn}
\affiliation{Institute of Modern Physics, Chinese Academy of Sciences, Lanzhou 730000,
China} \affiliation{University of Chinese Academy of Sciences, Beijing
100049, China}
\affiliation{Research Center for Hadron and CSR Physics, Institute of Modern Physics of
CAS and Lanzhou University, Lanzhou 730000, China}
\author{Jun He}
\thanks{Corresponding author: junhe@impcas.ac.cn}
\affiliation{Institute of Modern Physics, Chinese Academy of Sciences, Lanzhou 730000,
China}
\affiliation{Research Center for Hadron and CSR Physics, Institute of Modern Physics of
CAS and Lanzhou University, Lanzhou 730000, China}
\affiliation{State Key Laboratory of Theoretical Physics, Institute of Theoretical
Physics, Chinese Academy of Sciences, Beijing 100190, China}
\begin{abstract}
Inspired by the preliminary experimental data by the CLAS Collaboration, the
$K^{\ast 0}\Lambda$ photoproduction off a neutron is studied within an
effective Lagrangian approach. The contributions from the Born terms
including $s$, $u$, and $t$ channels are considered to calculate the
amplitude, with which the cross sections are calculated and compared with
preliminary CLAS data. The theoretical results indicate that the
contribution from the $t$-channel $K$ exchange plays a dominant role for the
$K^{\ast 0}\Lambda$ photoproduction. The contribution from the $\kappa$
exchange is found one order of magnitude smaller than that from the $K$
exchange. Both the Regge model and the Feynman model are applied to treat
the $t$-channel contribution. The discrepancy between two models is found
small in the energy range of CLAS data and predicted to become obvious at
energies higher than 3 GeV. More precise experimental data especially at
backward angles will be helpful to further understand the interaction
mechanism of the $K^{\ast 0}\Lambda$ photoproduction.
\end{abstract}

\maketitle

\section{Introduction}

In recent years, the strange meson photoproductions are widely investigated
in both experiment and theory. In addition to the photoproductions of ground-state
strange meson and strange baryon, i.e., $K\Lambda $ and $K\Sigma $, the
photoproductions of an excited meson or baryon have been of interest in
hadron physics. Among these processes, the photoproductions of the $\Lambda
(1520)$ and the $\Sigma (1385)$ were found important to study the nucleon
resonances especially those around 2 GeV~\cite%
{Nam:2005uq,Xie:2010yk,He:2012ud,He:2013ksa,He:2014gga,Wang:2015}. The $%
K^{\ast }$ photoproduction was also suggested to be useful to study the
nucleon resonances and the $\kappa $ meson~\cite%
{Oh:2006hm,Oh:2006in,Ozaki:2009wp,Kim:2011rm,Kim:2014hha}.

The $\gamma n\rightarrow K^{\ast 0}\Lambda $ reaction attracts special
attention because of its absence of the contact term, which is usually dominant in a
photoproduction process~\cite%
{Nam:2005uq,Xie:2010yk,He:2012ud,He:2013ksa,He:2014gga,Wang:2015}. The
interaction mechanism is illustrated in Fig. 1.
\begin{figure}[tbph]
\begin{center}
\includegraphics[scale=0.7]{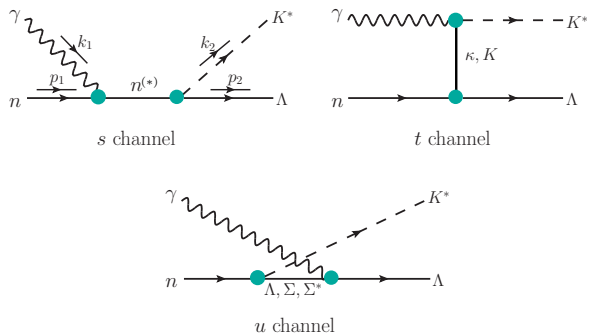}
\end{center}
\caption{(Color online) Feynman diagrams for the $\protect\gamma %
n\rightarrow K^{\ast 0}\Lambda $ reaction.}
\label{Fig:fyd}
\end{figure}

These include the $s$ channel with nucleon and its resonances, $t$ channels with $K$ and $%
\kappa $ exchanges, and the $\ u$ channel with hyperon ($\Lambda $, $\Sigma $ and $%
\Sigma ^{\ast }$). Usually, the contact term should be included to keep the
gauge invariance of the contributions from the $s$, $u$, and $t$ channels~\cite%
{Haberzettl:1997jg,Haberzettl:2015exa}. However, as shown in Refs.~\cite%
{Oh:2006hm,Kim:2011rm,Kim:2014hha} and this work, the amplitudes for the $%
\gamma n\rightarrow K^{\ast 0}\Lambda $ reaction from these channels are
gauge invariant themselves, which leads to the absence of the contact term. The $%
t$-channel $K^{\ast }$ exchange vanishes also because of the neutral charge of
final $K^{\ast }$. Usually, the contribution from the $s$ channel with the  nucleon
pole is very small and negligible and the  $u$-channel contribution is
important only at backward angles~\cite{He:2013ksa,He:2014gga}. Hence, the $%
\gamma n\rightarrow K^{\ast 0}\Lambda $ reaction is a very ideal channel to
investigate the nucleon resonances and the $t$-channel $K$ and $\kappa$
exchanges.

Thanks to the experimental data released from the facilities such as CLAS~%
\cite{Tang:2013gsa}, the $K^*\Lambda$ photoproduction off a proton was
widely investigated theoretically~\cite%
{Oh:2006hm,Oh:2006in,Ozaki:2009wp,Kim:2011rm,Kim:2014hha}. However, because of a
lack of experimental data,  study of the $K^{\ast 0}\Lambda$ photoproduction off a
neutron is scarce except for some predictions~\cite%
{Oh:2006hm,Kim:2011rm,Kim:2014hha}. Recently, {the} CLAS Collaboration
reported preliminary experiment{al} data for the $\gamma n\rightarrow
K^{\ast 0}\Lambda $ reaction \cite{Mattione:2014xba}. In Refs.~\cite%
{Oh:2006hm,Kim:2014hha}, the contribution from the nucleon resonance was
suggested to be small and the $\kappa$ exchange is also suppressed. The CLAS data
provide an opportunity to do a preliminary check of these opinions. And if
they are right, the $K$ exchange becomes dominant at this interaction, which
is helpful for clarifying the role of the $t$-channel contribution in the $%
\gamma n\rightarrow K^{\ast 0}\Lambda $ reaction.

In this work, within an effective Lagrangian approach, we analyze the $%
\gamma n\rightarrow K^{\ast 0}\Lambda $ reaction based on the preliminary
CLAS data. The interaction mechanism illustrated in Fig. 1 will be included
to calculate the differential cross section. In Ref.~\cite{Ozaki:2009wp} the
Regge model is found essential to reproduce the experimental data of the charged
$K^*$ photoproduction. So, in this work, both the Regge model and the
Feynman model will be applied to treat the $t$-channel contribution. The
nucleon resonances are not included in the calculation but a discussion will be
provided.

This paper is organized as follows. After introduction, formalism including
Lagrangians and amplitudes of the $\gamma n\rightarrow K^{\ast 0}\Lambda $
reaction are presented in Sec. II. The numerical results of the differential
cross section follow in Sec. III and are compared with CLAS data. Finally,
the paper{\ concludes} with the summary and discussion.

\section{Formalism}

To obtain the amplitude from the interaction mechanisms in Fig. 1 to
calculate the observables, we need the relevant Lagrangians.

\subsection{Lagrangian}

For the $s$ channel, the involved Lagrangians read as \cite%
{Oh:2006hm,Oh:2006in,Ozaki:2009wp,Kim:2011rm,Kim:2014hha},
\begin{align}
\mathcal{L}_{\gamma NN}& =-e\bar{N}(Q_{N}\rlap{$\slash$}A-\frac{\kappa _{N}}{%
4m_{N}}\sigma ^{\mu \nu }F^{\mu \nu })N~, \\
\mathcal{L}_{K^{\ast }N\Lambda }& =-g_{K^{\ast }N\Lambda }\bar{N}\Lambda
\left( \rlap{$\slash$}K^{\ast }-\frac{\kappa _{K^{\ast }N\Lambda }}{2m_{N}}%
\sigma _{\mu \nu }\partial ^{\nu }K^{\ast \mu }\right) +\text{h.c.}~,
\end{align}%
with the isodoublets
\begin{equation}
K^{T}=\binom{K^{+}}{K^{0}},\text{ }K^{\ast }=\binom{K^{\ast +}}{K^{\ast 0}}%
,\quad N=\binom{p}{n},
\end{equation}%
where $F^{\mu \nu }=\partial ^{\mu }A^{\nu }-\partial ^{\nu }A^{\mu }$, and $%
A^{\mu }$, $K$, $K^{\ast \mu }$ and $N$ are the photon, kaon, $K^{\ast }$
and nucleon fields. $Q_{N}$ is the charge of the nucleon in the unit of $e=\sqrt{%
4\pi \alpha }$ with $\alpha $ being the fine-structure constant{, and} $%
\kappa _{N}=-1.913$ {is anomalous magnetic moment} of the neutron. We adopt
coupling constants $g_{K^{\ast }N\Lambda }=-4.26$ GeV$^{-1}$ and $\kappa
_{K^{\ast }N\Lambda }=2.66$ GeV$^{-1}$ determined by the Nijmegen potential~%
\cite{Stoks:1999bz}.

For the {$t$-channel }$K$ exchange, we need following Lagrangians:
\begin{eqnarray}
\mathcal{L}_{\gamma KK^{\ast }} &=&g_{\gamma KK^{\ast }}\epsilon ^{\mu \nu
\alpha \beta }(\partial _{\mu }A_{\nu })(\partial _{\alpha }K_{\beta }^{\ast
})K+\text{h.c.}~, \\
\mathcal{L}_{KN\Lambda } &=&-ig_{KN\Lambda }\bar{N}\gamma _{5}\Lambda K+%
\text{h.c.}.
\end{eqnarray}
The photon coupling $g_{\gamma KK^{\ast }}$ is determined by the $K^{\ast }$
radiative decay width \cite{Agashe:2014kda}, which leads to a value $%
g_{\gamma KK^{\ast }}=-0.388$ GeV$^{-1}$ for netural $K^{\ast }$. Moreover,
coupling constant $g_{KN\Lambda }$ can be determined by using the SU(3) flavor
symmetry relation~\cite%
{Oh:2006hm,Oh:2006in,Ozaki:2009wp,Kim:2011rm,Kim:2014hha},
\begin{equation}
g_{KN\Lambda }=-\frac{1}{\sqrt{3}}(1+2\alpha )g_{\pi NN}=-13.24,
\end{equation}%
with $\alpha =0.365$ and $g_{\pi NN}^{2}/4\pi =14.0$.

The scalar $\kappa $ meson exchange is allowed in the $\gamma n\rightarrow
K^{\ast 0}\Lambda $ reaction, and there does not exist the contact term
which is dominant in the $K^*$ photoproduction off the proton~\cite%
{Ozaki:2009wp,Kim:2014hha}. It was suggested to investigate the role of the {$t$%
-channel }$\kappa $ exchange in the $K^{\ast }$ photoproduction while
the result in Ref.~\cite{Oh:2006hm} suggested its contribution should be
suppressed. In this work, we will consider $\kappa$ exchange in the $t$ channel.
To study scalar $\kappa $ meson exchange, we need to construct the effective
Lagrangians for $\gamma K^{\ast }\kappa $ and $\kappa N\Lambda $ couplings~%
\cite{Oh:2006hm,Oh:2006in,Ozaki:2009wp,Kim:2011rm,Kim:2014hha},
\begin{eqnarray}
\mathcal{L}_{\gamma K^{\ast }\kappa } &=&g_{\gamma K^{\ast }\kappa }A^{\mu
\nu }\bar{\kappa}K_{\mu \nu }^{\ast }+\text{h.c.}~, \\
\mathcal{L}_{\kappa N\Lambda } &=&-g_{\kappa N\Lambda }\bar{N}\kappa \Lambda
+\text{h.c.}~,
\end{eqnarray}%
with $A^{\mu \nu }=\partial _{\mu }A_{\nu }-\partial _{\nu }A_{\mu }$ and%
\begin{equation}
\kappa =(\kappa ^{+},\kappa ^{0}),\text{ }\bar{\kappa}^T=({\kappa ^{-}},{%
\bar{\kappa}^{0}})^T.
\end{equation}%
Usually, the mass and width of the $\kappa $ meson are $m_{\kappa }=700\sim 900$
MeV and $\Gamma _{\kappa }=400\sim 770$ MeV, respectively ~\cite%
{Agashe:2014kda}. Here we employ $m_{\kappa }=800$ MeV, $\Gamma _{\kappa
}=550$ MeV, and $g_{\gamma K^{\ast }\kappa }g_{\kappa N\Lambda }=-2.2e$ GeV$%
^{-1}$ for neutral $\kappa $ meson~\cite{Oh:2006hm,Oh:2006in}.

For the $u$ channel from the $\Lambda (1116), \Sigma (1193)$ and $\Sigma ^{\ast
}(1385) $ exchanges, the effective Lagrangians depicting $\gamma \Lambda
\Lambda $, $\gamma \Sigma \Lambda $, $K^{\ast }N\Sigma $, $KN\Sigma ^{\ast }$
and $\gamma \Lambda \Sigma ^{\ast }$ couplings are of forms~\cite%
{Oh:2006hm,Oh:2006in,Ozaki:2009wp,Kim:2011rm,Kim:2014hha},
\begin{align}
\mathcal{L}_{\gamma \Lambda \Lambda }& =\frac{e\kappa _{\Lambda }}{%
2m_{\Lambda }}\bar{\Lambda}\sigma _{\mu \nu }\partial ^{\nu }A^{\mu }\Lambda
\text{,} \\
\mathcal{L}_{\gamma \Sigma \Lambda }& =\frac{e\mu _{\Sigma \Lambda }}{%
2m_{\Lambda }}\bar{\Sigma}^{0}\sigma _{\mu \nu }\partial ^{\nu }A^{\mu
}\Lambda +\text{h.c.}~, \\
\mathcal{L}_{K^{\ast }N\Sigma }& =-g_{K^{\ast }N\Sigma }\bar{N}\Sigma \left( %
\rlap{$\slash$}K^{\ast }-\frac{\kappa _{K^{\ast }N\Sigma }}{2m_{N}}\sigma
_{\mu \nu }\partial ^{\nu }K^{\ast \mu }\right) +\text{h.c.}~, \\
\mathcal{L}_{K^{\ast }N\Sigma ^{\ast }}& =-i\frac{f_{K^{\ast }N\Sigma ^{\ast
}}^{(1)}}{m_{K^{\ast }}}\overline{K}_{\mu \nu }^{\ast }\bar{\Sigma}^{\ast
\mu }\cdot \tau \gamma ^{\nu }\gamma _{5}N  \notag \\
& \text{ \ \ }-\frac{f_{K^{\ast }N\Sigma ^{\ast }}^{(2)}}{m_{K^{\ast }}}%
\overline{K}_{\mu \nu }^{\ast }\bar{\Sigma}^{\ast \mu }\cdot \tau \gamma
_{5}\partial ^{\nu }N  \notag \\
& \text{ \ \ }-\frac{f_{K^{\ast }N\Sigma ^{\ast }}^{(3)}}{m_{K^{\ast }}}%
\partial ^{\nu }\overline{K}_{\mu \nu }^{\ast }\bar{\Sigma}^{\ast \mu }\cdot
\tau \gamma _{5}N+\text{h.c.}~, \\
\mathcal{L}_{\gamma \Lambda \Sigma ^{\ast }}& =\frac{ieg_{1}}{2m_{\Lambda }}%
\bar{\Lambda}\gamma _{\nu }\gamma _{5}F^{\mu \nu }\Sigma _{\mu }^{\ast }
\notag \\
& \text{ \ \ }-\frac{eg_{2}}{(2m_{\Lambda })^{2}}\partial _{\nu }\bar{\Lambda%
}\gamma _{5}F^{\mu \nu }\Sigma _{\mu }^{\ast }+\text{h.c.}~,
\end{align}%
where $\kappa _{\Lambda }=-0.61$, $\mu _{\Sigma \Lambda }=1.62$, $g_{K^{\ast
}N\Sigma }=-2.46$ and $\kappa _{K^{\ast }N\Sigma }=-0.47$ are adopted \cite%
{Oh:2006hm,Oh:2006in}. The coupling $f_{K^{\ast }N\Sigma ^{\ast }}^{(1)}$
can be obtained from SU(3) flavor symmetry relations, which gives $%
f_{K^{\ast }N\Sigma ^{\ast }}^{(1)}=-2.6$~\cite{Oh:2006in}. In this work, we
take $f_{K^{\ast }N\Sigma ^{\ast }}^{(2)}=f_{K^{\ast }N\Sigma ^{\ast
}}^{(3)}=0$ because of the lack of relevant information ~\cite{Oh:2006in}. The
electric magnetic coupling constants $g_{1}$ and $g_{2}$ are determined from
the decay width $\Gamma _{\Sigma ^{\ast }\rightarrow \Lambda \gamma }$ as
well as magnetic dipole ($M1$) and electric quadrupole ($E2$) moments, which
leads to values $(g_{1},g_{2})=(3.78,3.18)$~ \cite{Agashe:2014kda}.

{In our calculation, the phenomenological form factors are introduced to
account for the internal structure of hadrons.} We adopt the{\ functional
form} used in Refs.~\cite{Oh:2006hm,Oh:2006in},
\begin{equation}
\mathcal{F}_{s/u}(q_{ex}^{2})=\frac{\Lambda _{s/u}^{4}}{\Lambda
_{s/u}^{4}+(q_{ex}^{2}-m_{ex}^{2})^{2}}~,
\end{equation}%
for the $s$ and $u$ channels, and
\begin{equation}
\mathcal{F}_{t}(q_{ex}^{2})=\frac{\Lambda _{t}^{2}-m_{ex}^{2}}{\Lambda
_{t}^{2}-q_{ex}^{2}}~,
\end{equation}%
for the $t$ channel. Here $q_{ex}$ and $m_{ex}$ are four-momentum and mass of
the exchanged hadron, respectively. The values of cutoffs $\Lambda _{s}$, $%
\Lambda _{u}$ and $\Lambda _{t}$ will be{\ determined by fitting
experimental data}.

\subsection{Amplitudes for $\protect\gamma n\rightarrow K^{\ast 0}\Lambda $
process}

After above preparations, the invariant scattering amplitudes for the $%
\gamma n\rightarrow K^{\ast 0}\Lambda $ reaction{\ can be written as},
\begin{equation}
-i\mathcal{M}_{i}=\epsilon^{\nu*}_{\lambda _{K^{\ast }}}(k_{2})~\bar{u}%
_{\lambda _{\Lambda }}(p_{2})~A_{i,~\mu \nu }~u_{\lambda
_{n}}(p_{1})~\epsilon ^{\mu }_{\lambda _{\gamma }}(k_{1}).
\end{equation}%
Here $u_{\lambda _{(\Lambda,n) }}$ is the Dirac spinor of neutron or $\Lambda
$ with helicity $\lambda _{(\Lambda,n) }$.  $\epsilon_{\lambda_{(K^*,%
\gamma)}}$ is the polarization vector of $K^{\ast }$ or photon with helicity
$\lambda _{(K^*,\gamma) }$.

The reduced amplitudes $A_{i}^{\mu \nu }$ for $s$-, $t$- and $u$-channel
contributions read,
\begin{align}
A_{s(N)}^{\mu \nu }& =\frac{eg_{K^{\ast }N\Lambda }}{2m_{N}}\frac{\kappa _{N}%
}{s-m_{N}^{2}}\left( \gamma ^{\nu }-\frac{\kappa _{K^{\ast }N\Lambda }}{%
2m_{N}}\gamma ^{\nu }\rlap{$\slash$}k_{2}\right)  \notag \\
& \mbox{}\qquad \times (\rlap{$\slash$}k_{1}+\rlap{$\slash$}%
p_{1}+m_{N})\gamma ^{\mu }\rlap{$\slash$}k_{1}\mathcal{F}_{s}~, \\
A_{t(K)}^{\mu \nu }& =\frac{-ig_{\gamma KK^{\ast }}g_{KN\Lambda }}{%
t-m_{K}^{2}}\epsilon ^{\mu \nu \alpha \beta }k_{1\alpha }k_{2\beta }\gamma
_{5}\mathcal{F}_{t}~,  \label{Eq: tchannel} \\
A_{t(\kappa )}^{\mu \nu }& =\frac{-2g_{\gamma K^{\ast }\kappa }g_{\kappa
N\Lambda }}{t-\left( m_{\kappa }-i\Gamma _{\kappa }/2\right) ^{2}}\left(
k_{1}k_{2}g^{\mu \nu }-k_{1}^{\nu }k_{2}^{\mu }\right) \mathcal{F}_{t}~, \\
A_{u(\Lambda )}^{\mu \nu }& =\frac{e\kappa _{\Lambda }}{2m_{\Lambda }}\frac{%
g_{K^{\ast }N\Lambda }}{u-m_{\Lambda }^{2}}\gamma ^{\mu }\rlap{$\slash$}%
k_{1}\left( \rlap{$\slash$}p_{1}-\rlap{$\slash$}k_{2}+m_{\Lambda }\right)
\notag \\
& \mbox{}\qquad \times \left( \gamma ^{\nu }-\frac{\kappa _{K^{\ast
}N\Lambda }}{2m_{N}}\gamma ^{\nu }\rlap{$\slash$}k_{2}\right) \mathcal{F}_{u}%
\text{ }, \\
A_{u,(\Sigma )}^{\mu \nu }& =\frac{-e\mu _{\Sigma \Lambda }}{2m_{\Lambda }}%
\frac{g_{K^{\ast }N\Sigma }}{u-m_{\Sigma }^{2}}\gamma ^{\mu }\rlap{$\slash$}%
k_{1}\left( \rlap{$\slash$}p_{1}-\rlap{$\slash$}k_{2}+m_{\Sigma }\right)
\notag \\
& \mbox{}\qquad \times \left( \gamma ^{\nu }-\frac{\kappa _{K^{\ast }N\Sigma
}}{2m_{N}}\gamma ^{\nu }\rlap{$\slash$}k_{2}\right) \mathcal{F}_{u}\text{ },
\\
A_{u(\Sigma ^{\ast })}^{\mu \nu }& =\frac{-ef_{K^{\ast }N\Sigma ^{\ast
}}^{(1)}}{m_{K^{\ast }}} \left( \frac{g_{1}}{2m_{\Lambda }}\gamma _{\nu
}\gamma _{5}+\frac{g_{2}}{4m_{\Lambda }^{2}}p_{2\nu }\gamma _{5}\right)
\notag \\
& \mbox{}\qquad \times \left( k_{1}^{\beta }g^{\mu \nu }-k_{1}^{\nu }g^{\mu
\beta }\right)\frac{\rlap{$\slash$}q_{u}+m_{\Sigma ^{\ast }}}{u-m_{\Sigma
^{\ast }}^{2}} G_{\beta \alpha }\gamma _{\sigma }\gamma _{5}  \notag \\
& \mbox{}\qquad \times \left( k_{2}^{\alpha }g^{\nu \sigma }-k_{2}^{\sigma
}g^{\nu \alpha }\right) \mathcal{F}_{u}\text{ },
\end{align}%
with%
\begin{equation}
G_{\beta \alpha }=g_{\beta \alpha }-\frac{1}{3}\gamma _{\beta }\gamma
_{\alpha }-\frac{2(q_{u})_{\beta }(q_{u})_{\alpha }}{3m_{\Sigma ^{\ast }}^{2}%
} -\frac{\gamma _{\beta }(q_{u})_{\alpha }-\gamma _{\alpha }(q_{u})_{\beta }%
}{3m_{\Sigma ^{\ast }}},
\end{equation}%
where $s=q_{s}^{2}=(k_{1}+p_{1})^{2}$, $t=q_{t}^{2}=(k_{1}-k_{2})^{2}$ and $%
u=q_{u}^{2}=(p_{2}-k_{1})^{2}$ are the Mandelstam variables.

In Ref.~\cite{Nam:2010au} and our previous works~\cite%
{He:2012ud,He:2013ksa,He:2014gga,Wang:2015}, an interpolating Regge
treatment was introduced to interpolate the Regge trajectories smoothly to the
Feynman propagator at low energy at a cost of four additional free
parameters introduced. Since there are only 17 data points in preliminary
CLAS data, we do not adopt the interpolating Reggiezed treatment but the
Feynman model and the Regge model. For the Feynman model, the $t$-channel
amplitude in Eq.~(\ref{Eq: tchannel}) is applied directly. The Regge model
can be introduced by replacing the $t$-channel Feynman propagator with the
Regge propagator as follows \cite{Titov:2005kf,Guidal:1997hy,Corthals:2006nz}%
,
\begin{eqnarray}
\frac{1}{t-m_{K}^{2}} &\rightarrow &(\frac{s}{s_{\text{scale}}})^{\alpha
_{K}(t)}\frac{\pi \alpha _{K}^{\prime }}{\Gamma \lbrack 1+\alpha
_{K}(t)]\sin [\pi \alpha _{K}(t)]}, \\
\frac{1}{t-m_{\kappa }^{2}} &\rightarrow &(\frac{s}{s_{\text{scale}}}%
)^{\alpha _{\kappa }(t)}\frac{\pi \alpha _{\kappa }^{\prime }}{\Gamma
\lbrack 1+\alpha _{\kappa }(t)]\sin [\pi \alpha _{\kappa }(t)]}.
\end{eqnarray}%
The $\alpha _{K}^{\prime }$ is the slope of the trajectory and the scale
factor $s_{\text{scale}}$ is fixed at 1 GeV$^{2}$. In addition, the kaonic
Regge trajectories $\alpha _{K}(t)$ and $\alpha _{\kappa }(t)$ read as \cite%
{Ozaki:2009wp},%
\begin{equation*}
\alpha _{K}(t)=0.7\mathrm{GeV}^{-2}(t-m_{K}^{2}),\ \alpha _{\kappa }(t)=0.7%
\mathrm{GeV}^{-2}(t-m_{\kappa }^{2}).\quad \ \
\end{equation*}

\section{Numerical results}

As mentioned above, the $s$ channel with nucleon pole, the $t$ channels with
$K$ and $\kappa $ exchanges, and the $u$ channels with $\Lambda ,\Sigma $ and $%
\Sigma ^{\ast }$ exchanges are included to calculate its differential cross
section, which will be compared with the CLAS data. The differential cross
section in the center of mass (c.m.) frame is given by
\begin{equation}
\frac{d\sigma }{d\cos \theta }=\frac{1}{32\pi s}\frac{\left\vert \vec{k}%
_{2}^{{~\mathrm{c.m.}}}\right\vert }{\left\vert \vec{k}_{1}^{{~\mathrm{c.m.}}%
}\right\vert }\left( \frac{1}{4}\sum\limits_{\lambda }\left\vert \mathcal{M}%
\right\vert ^{2}\right),
\end{equation}%
where $s=(k_{1}+p_{1})^{2}$, and $\theta $ denotes the angle of the outgoing
$K^{\ast 0}$ meson relative to the beam direction in the c.m. frame. $\vec{k}%
_{1}^{{~\mathrm{c.m.}}}$ and $\vec{k}_{2}^{{~\mathrm{c.m.}}}$ are the
three-momenta of the initial photon beam and final $K^{\ast}$, respectively.

\subsection{Fitting procedure}

The preliminary CLAS data~\cite{{Mattione:2014xba}} will be fitted with the
help of the \textsc{minuit} code in the \textsc{cernlib}. In literature~\cite%
{Oh:2006hm,Oh:2006in,Ozaki:2009wp,Kim:2011rm,Kim:2014hha}, the cutoff values
for the$s$ channel and $u $ channel for the $\gamma N\rightarrow K^{\ast
}\Lambda $ were usually determined to be 0.9 GeV. Numerical tests show that
for the present case the contribution from the $s$ channel is negligible and
contribution from the $u$ channel is important only at the backward angle where the
experimental data do not exist. Thus, in this work the cutoffs $\Lambda_{s}$
and $\Lambda_u$ are fixed as Refs.~\cite%
{Oh:2006hm,Oh:2006in,Ozaki:2009wp,Kim:2011rm,Kim:2014hha}. Other parameters
have been fixed as presented in the previous section, and we would like to note
that with these parameters the cross section of $K^{*+}\Lambda$
photoproduction was reproduced~\cite{Ozaki:2009wp}. Hence, a one-parameter $%
\chi ^{2}$ fitting ($\Lambda _{t}$) will be performed to fit the CLAS data
of the differential cross section $d\sigma /d\cos \theta $.

The CLAS experimental data of the differential cross section ${d\sigma }/{d\cos
\theta }$ include 17 data points at three intervals of the beam energy $%
E_\gamma=$1.9$-$2.1, 2.1$-$2.3 and 2.3$-$2.5 GeV. The data will be fitted in two
schemes, the Reggeized model and the Feynman model, as interpreted in
the previous section. The fitted values of the only one free parameter $\Lambda_t
$ in two schemes are listed in Table I.

\renewcommand\tabcolsep{0.52cm} \renewcommand{\arraystretch}{1.4}
\begin{table}[h!]
\caption{The fitted value of free parameter $\Lambda_t$ in the unit of GeV
and compared with values for the $K^{*+}\Lambda$ photoproduction~%
\protect\cite{Ozaki:2009wp}. }%
\begin{tabular}{lccc}
\hline
& $\Lambda _{t}$ & $\chi ^{2}/dof$ & $\Lambda _{t}$ in Ref.~\cite%
{Ozaki:2009wp} \\ \hline
Feynman & 1.05$\pm 0.01$ & 1.99 & 1.15 \\
Regge & 2.28$\pm 0.12$ & 1.68 & 1.55 \\ \hline
\end{tabular}%
\end{table}

One notices that the fitted cutoff values in the Feynman model and the Regge
model are close to the values for the $K^{*+}\Lambda$ photoproduction off the
proton target~\cite{Ozaki:2009wp}.

\subsection{\boldmath Cross section for the $\protect\gamma n\rightarrow
K^{\ast 0}\Lambda $ reaction}

The differential cross section ${d\sigma }/{\ d\cos \theta }$ obtained in
two schemes are illustrated in Fig. 2. Both results are acceptable
considering that only one parameter is fitted in the current work. The difference
between the best-fitted $\chi^2$ for two schemes are small with values $1.99$
and $1.68$ for the Feynman and the Regge models, respectively. The slope of curve in the full model for the Regge model is steeper than
that for the Feynman model, which results in  the Regge model working
better at forward angles while the Feynman model is better at medium angles ( $\cos\theta
$ around 0) .

\begin{figure}[htbp]
\centering
\includegraphics[scale=0.32]{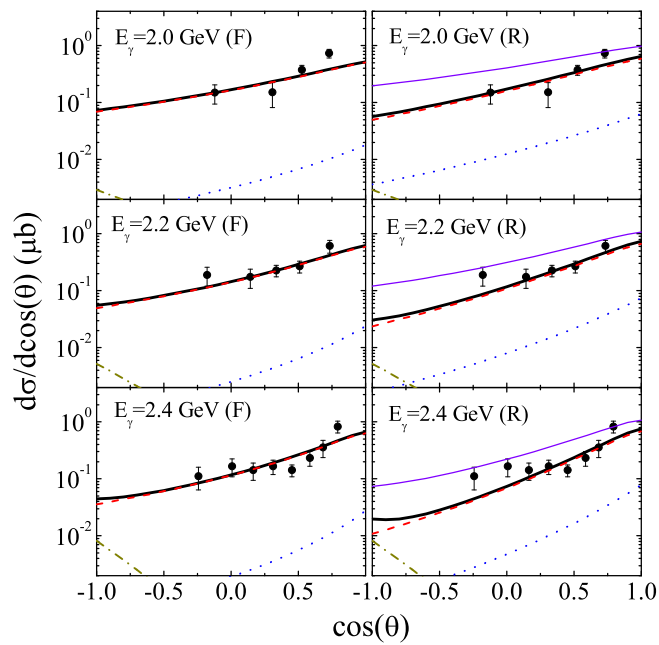}
\caption{(Color online) The differential cross section $d\protect\sigma %
/d\cos \protect\theta $ for the $K^{\ast 0}\Lambda$ photoproduction from the
neutron as a function of $\cos \protect\theta $. The experimental data are from Ref.
\protect\cite{Ozaki:2009wp}. The marks $(F)$ and $(R)$ are for the Feynman
model and the Regge model, respectively. The full (black), dashed (red),
dotted (blue), and dash-dotted (dark yellow) lines are for the full model, $K$
exchange, $\protect\kappa$ exchange, and $u$ channel. The thinner full line (violet) is for the Regge model without the  form factor.}
\end{figure}

One can find that the $K$ exchange is dominant at energies $E_\gamma=$
2.0$-$2.4 GeV. The $\kappa$-exchange contribution is much smaller than the
contribution from $K$ exchange and almost has no effect on the differential
cross sections in the full model. The $s$-channel contribution is negligible as
the other photoproduction process. The $u$ channel works at backward angles and
leads to a small increase of the differential cross section.

To provide a clearer picture of the interaction mechanism of the $K^{\ast
0}\Lambda$ photoproduction, more experimental data at higher energies are
expected. Here we present our prediction of the differential cross section
at higher energies upto 4 GeV in Fig. 3. Differences between the Regge and
the Feynman model at low energies are small but become large at higher
energies, which will be useful in clarifying the role of the Reggeized
treatment and can be tested by further experiment.

\begin{figure}[htbp]
\centering
\includegraphics[scale=0.32]{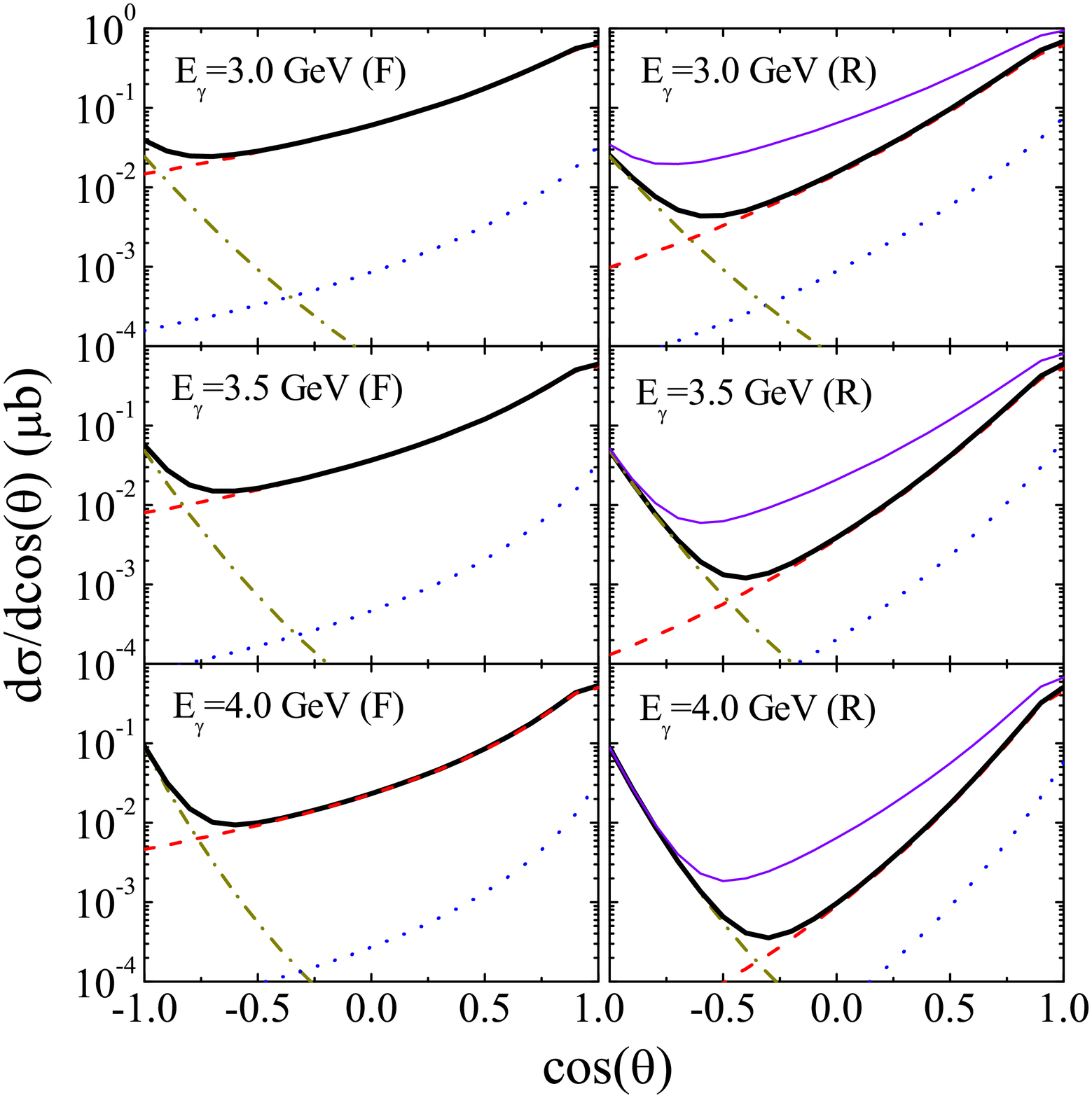}
\caption{(Color online) The differential cross section $d\protect\sigma %
/d\cos \protect\theta $ for the $K^{\ast 0}\Lambda$ photoproduction from the
neutron as a function of $\cos \protect\theta $. The notations are as in
Fig. 2.}
\end{figure}
The contribution from the $K$ exchange is dominant at forward angles while the
contribution from the $u$ channel becomes more and more important at backward
angles with the increase of the energy. The differential cross section in
full model is almost from these two contributions.

The formalism and parameters adopted in this work are the same as those in Ref. \cite{Ozaki:2009wp} except for cutoff $\Lambda$.  In Ref. \cite{Ozaki:2009wp} the charged $K^*$ photoproduction was studied and the experimental data were well reproduced by both the Feynman and the Regge models. It suggests such formalism is effective for both charged and neutral $K^*$ photoproducitons.

One may notice that following Ref. \cite{Ozaki:2009wp} the form factor for the $t$ channel is not removed after the Regge treatment  in the current work. In fact, because the Regge treatment is phenomenological, there does not exist any strict theoretical requirement to remove the form factor, or keep it. In this work, we also present the results without the form factor in Figs. 2 and 3.
After removing the form factor from the $t$-channel contribution, the differential cross sections increase at all energies considered, and their slopes become less steep. It is easy to understand because the form factor provides an increased suppression of the $t$-channel exchanges at backward angles. To reproduce the experimental data  after removing the form factor, we will have to decrease the coupling constant $g_{KN\Lambda}$ because the $K$ exchange is dominant and other parameters are fixed in the Regge model.

The total cross section of the $\gamma n\rightarrow K^{\ast 0}\Lambda $
process is illustrated in Fig. 4. It is found that the $t$-channel $K$
exchange is dominant at energies from the threshold up to 4.5 GeV. The
contribution from the $u$ channel increases with the increase of energy. The
contribution from the $t$-channel $\kappa $ exchange is larger but still
negligible when the Reggeized treatment is considered. At high energies, the
total cross section with the Regge model decreases more rapidly than that
with the Feynman model.

\begin{figure}[h!]
\begin{center}
\includegraphics[scale=0.33,clip]{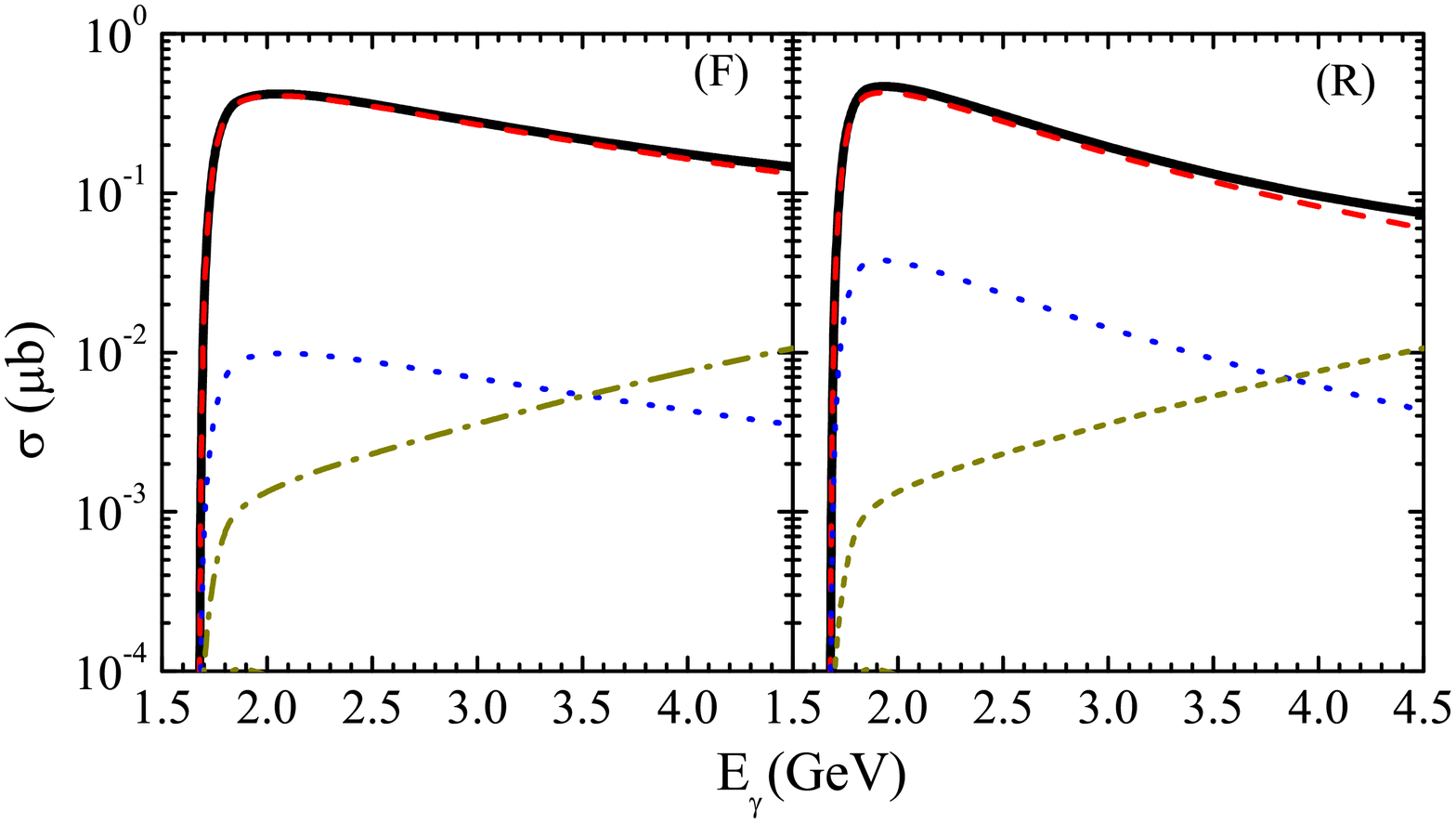}
\end{center}
\caption{(Color online) (a): Total cross section for $\protect\gamma %
n\rightarrow K^{\ast 0}\Lambda $ reaction. The notations are as in Fig. 2.}
\label{Fig:total}
\end{figure}

\section{Summary and discussion}

Within an effective Lagrangian approach, a one-parameter fitting is done to
preliminary CLAS data of the $K^{\ast }\Lambda$ photoproduction off the
neutron with the Feynman and the Regge models. The numerical results
indicate that the differential cross section is well reproduced by Regge
model with $\chi ^{2}/dof=1.68$. The Feynman model works better at medium
angles with a little worse $\chi^2$=1.99. The discrepancies between the
Regge model and the Feynman model are small at low energies $E_\gamma=$2.0$-$2.4
GeV but become obvious at higher energies especially at backward angles.

The result suggests that the $t$-channel $K$ exchange  is overwhelmingly
dominant at $K^{\ast }\Lambda$ photoproduction off the neutron. It was
suggested that the $K^*$ photoproduction is the ideal process to study the $%
\kappa$ exchange. However, our results confirmed the conclusion in Ref.~\cite%
{Oh:2006hm} that the contribution from $\kappa$ exchange is much smaller
than the contribution from the dominant $K$ exchange (by one order of magnitude),
which makes it difficult to study $\kappa$ meson in this process.

Nucleon resonances are not included in the fitting of the experimental data.
The small $\chi^2$ suggests their contribution should not be very large~\cite%
{Oh:2006hm,Kim:2014hha}. However, the nucleon resonances still may provide
a considerable contribution at medium and backward angles. It is interesting
to see that the Feynman model works better at forward angles while the Regge
model is better at medium angles, where the differential cross section is smaller than
those at forward angles (see Fig. 2). If we recall that the differential
cross section from a nucleon resonance decaying to $K^*\Lambda$ in the $S$ wave is
flat with variation of $\cos\theta$, it is possible that nucleon resonances,
such as $N(2120)$ which provides considerable contribution to the $\gamma
p\to K^{*+}\Lambda$ reaction, have an observable effect on the differential
cross section at backward angles for the $K^{\ast 0}\Lambda$
photoproduction. Another possibility is that the interpolating Regge treatment
in Ref.~\cite{Nam:2010au} should be introduced, with which the Regge
treatment works at forward angles and the Feynman propagator works at backward
angles. Because more free parameters will be introduced (there are only 17
data points in the preliminary CLAS experiment with considerable uncertainties
for data points at medium angles~\cite{Mattione:2014xba}), a fitting in the
interpolating Regge model and/or with nucleon resonance contribution is
impractical here.

The preliminary CLAS data are helpful to understand the interaction mechanism
of the $K^{*0}\Lambda$ photoproduction, such as confirming the dominance of the $%
K$ exchange and the smallness of nucleon resonance contribution. The precise
data at energies around 2 GeV and data at high energies, especially at
backward angles will be helpful to clarify the roles of nucleon resonances
and the Regge model, respectively.

\section{Acknowledgments}

We would like to thank Helmut Haberzettl for his useful and constructive
discussions about strange meson photoproductions. This project is supported
by the Major State Basic Research Development Program in China under Grant No.
2014CB845405 and the National Natural Science Foundation of China under
Grant No. 11275235.

\end{document}